\documentclass[twocolumn,pra,aps,raggedbottom,longbibliography]{revtex4-2}
\usepackage{times}

\usepackage{amsmath,amsfonts,amssymb,graphicx,hyperref,epstopdf,xcolor,tikz,scalerel,natbib,array,gensymb}
\usepackage{mathptmx,textcomp,braket}
\usepackage[T1]{fontenc}
\usepackage[utf8]{inputenc}
\usepackage{multirow}
\usepackage{array}

\definecolor{lime}{HTML}{A6CE39}
\DeclareRobustCommand{\orcidicon}
{
	\begin{tikzpicture} 
	\draw[lime, fill=lime] (0,0) circle [radius=0.15] node[white] {{\fontfamily{qag}\selectfont \tiny ID}};
	\draw[white, fill=white] (-0.0625,0.095) 	circle [radius=0.007];
	\end{tikzpicture}
	\hspace{-2.2mm}
}
\newcommand\orcidID[1]{\href{https://orcid.org/#1}{\orcidicon}}

\newcommand{\be}{\begin {equation}}
\newcommand{\ee}{\end {equation}}
\newcommand{\beqa}{\begin {eqnarray}}
\newcommand{\eeqa}{\end {eqnarray}}
\newcommand{\mb}{\mathbf}
\newcommand{\Sch}{Schr\"odinger }

\hypersetup{colorlinks,citecolor=blue,filecolor=black,linkcolor=blue,urlcolor=blue}

\begin{document}

\title{Bi-chromatic driver enabled control of high harmonic generation in atomic targets}

\author{Rambabu Rajpoot\orcidID{0000-0002-2196-6133}}
\email[E-mail: ]{ramrajpoot3@gmail.com}
\author{Amol R. Holkundkar\orcidID{0000-0003-3889-0910}}
\email[E-mail: ]{amol@holkundkar.in}
  

\affiliation{Department of Physics, Birla Institute of Technology and Science - Pilani, Rajasthan,
333031, India}

\date{\today}

\begin{abstract}
We investigated the high-order harmonic generation by interacting time-delayed, linearly polarized bi-chromatic laser pulses with the atomic target. The frequency ratio of secondary to primary fields ($\omega_2/\omega_1$), along with the relative carrier-envelope-phase (CEP) and respective time-delay between the two pulses, are very instrumental in controlling the ionization of the electron and so the quantum dynamics in continuum. We observed an optimum CEP ($\phi$) and time delay ($t_d$) for a given bi-chromatic frequency ratio, giving maximum cutoff energy. We studied the effect of CEP and time delay for fixed frequency ratio and observed that the harmonic cutoff could be controlled in an experimentally feasible way. Moreover, the harmonic yield in the energy range 140 eV $-$ 240 eV is also calculated, and it is observed that the harmonic yield scales as $\propto -\sqrt{\phi}$. The attosecond pulses for different CEP is also calculated by superposing the harmonics in the same energy range, and $\sim 200$ as pulses are observed with peak intensity decreasing with CEP, following the variation of respective harmonic yield with CEP. 
 
\end{abstract}

\maketitle

\section{Introduction}

The last couple of decades witnessed tremendous advancement in the study of laser-atom interaction and subsequent high-order harmonic generation (HHG) \cite{PhysRevLett.70.1599,Corkum1993_PRL}. The HHG is very crucial for the development of extreme ultraviolet (XUV) and soft x-ray radiation sources \cite{Liu2019SpecLett, Mairesse2003_Science} at an attosecond time scale \cite{Krausz2009_RMP,Hentschel2001_Nature,Corkum2007_NatPhy}, promising vast number of applications \cite{Chini2014_nat,Krausz2009_RMP,Heuser2016_PRA,Ayuso2018_JPhysB,Baykusheva2016_PRL,Reich2016_PRL}. The research in this field is also focused toward polarization control of the emitted harmonics \cite{Huo2021_PRA,Zhang2017_OptLett,Rajpoot2021_JPhysB}, and also extending the cutoff energy of the HHG and increasing the corresponding intensity of the emitted high-order harmonics \cite{Astiaso2016,Rajpoot_2020}.

The semi-classical \textit{three-step model} is instrumental for the basic understanding of the HHG. The HHG is described as a three-step process: ionization of the electron through tunneling, free propagation of electron in driving laser field, and finally, the recombination of the electron with the parent ion. The additional kinetic energy gained by the electron in its excursion is emitted in the form of the higher harmonics of the fundamental frequency of the interacting laser pulse. The free propagation of the ionized electron in the continuum presents an opportunity to regulate the electron path by modulating the waveform of the driving field \cite{Khodabandeh_2021,Shao2020_OEx, Greening2020_OEx}. An elegant and experimentally viable method to modulate the shape of the laser field is superposing a different frequency field to the existing one \cite{Bruner_2021,Katalin2015_IEEE,Jin_2016,Siegel2010}. The parameters like intensity, frequency and carrier envelope phase (CEP) of the new field along with the delay between the two pulses offers direct control over the shape of the driving field \cite{Chen2011, PhysRevLett.104.233901,PhysRevA.95.033413,PhysRevA.91.063408,PhysRevA.89.023431,Schutte2015}. Consequently, the properties of the associated high harmonic radiation and so the generated attosecond pulse can be controlled \cite{Greening2020,Bruner2018,PhysRevA.79.033842}.

Combining multi-color fields to sculpt a synthesized driver to control the quantum dynamics of the electron is a very well-known procedure. However, the ratio of the participating colors, along with the CEP of the same play a paramount role in determining the harmonic emission. In these types of setups, generally, a weaker field is taken to be of higher frequency than the fundamental frequency \cite{PhysRevA.91.063408, PhysRevA.89.023431}. In this work, we explored the effect of the frequency ratio $\omega_2/\omega_1$ on the harmonic emission process. We observed that the frequency ratio, the relative CEP of the two colors, and the delay between the two colors are equally responsible for the enhanced harmonic emission. For simplicity and to restrict the parameter space, we consider the field ratio of primary to secondary pulse to be $E_1/E_2 = 5$. However, we explored different $\omega_2/\omega_1$ ratios not restricted to $\omega_1 < \omega_2$. We focus on the time delay and CEP parameters, which are simple to regulate in experimental scenarios \cite{Kim2005_PhysRevLett} and studied the high-order harmonic generation by using the bi-chromatic field composed of time-delayed, linearly polarized, few-cycle laser pulses. The time-dependent \Sch equation (TDSE) is solved through the time-dependent generalized pseudo-spectral (TDGPS) method. The field profiles are considered such that the introduction of the delays does not affect the CEP of the resultant pulse. Details of the numerical methods are discussed in Sec. \ref{sec2}, followed by the results and the discussions in Sec. \ref{sec3}, by first exploring the CEP and delay effect for $\omega_2/\omega_1 = 0.5$ and later exploring the effect of CEP and time delay on HHG with different frequency ratios. The attosecond pulses are also constructed, where the peak intensity of the pulses can be controlled through CEP. The concluding remarks and future directions are discussed in Sec. \ref{sec4}. 
  
\section{Numerical Methods}
\label{sec2} 

We study the interaction of the linearly polarized laser pulse with a He atom by numerically solving the TDSE under single-active-electron (SAE) approximation using the TDGPS method \cite{TONG1997119}. The interaction of the linearly polarized laser ($m= 0$) with the spherically symmetric initial state of the He atom (1s state) will not alter the temporal evolution of the wavefunction in the azimuthal direction, and so effectively, it would suffice to solve the TDSE in radial and polar coordinates only. The TDSE in the length gauge is written as [atomic units are used throughout the manuscript] :
\be
i \frac{\partial}{\partial t} \psi(\mb{r},t) = \Big[ -\frac{\nabla^2}{2} + V(r) + \mb{r}\cdot\mb{E}(t) \Big] \psi(\mb{r},t),
\ee  
where $\mb{E}(t)$ is the temporal profile of the bi-chromatic driving laser electric field under dipole approximation. The driving field is obtained by combining two linearly polarized few-cycle laser pulses of frequency $\omega_1$ and $\omega_2$, respectively, and is described as (linearly polarized along $z$-direction):
\begin{multline}
	E(t) = E_1 f(t) \sin(\omega_1 t) + E_2 f(t-t_d) \sin[\omega_2 (t-t_d) + \phi],
	\label{laser_field}
\end{multline}
wherein $E_1$ and $E_2$ are the peak amplitudes of the electric fields of the two combining laser pulses. The pulse envelope $f(t) = \sin^2(\pi t / \tau)$ with total duration $\tau = 4T$, where $T = 2\pi/\omega_1$ is the period of the primary laser pulse. The quantities $t_d$ and $\phi$ are the time delay and relative carrier-envelope phase (CEP) between $\omega_1$ and $\omega_2$ fields, respectively. For all the cases presented in this work, we have considered the field amplitude $E_1 = 5E_2 \approx 0.17$ a.u., corresponding to the intensity $I = 10^{15}$ W cm$^{-2}$. Throughout this work, the frequency of the primary field $\omega_1 \sim 0.05695$ a.u. [800 nm laser] is considered, however $\omega_2$ is decided by the frequency ratio $\omega_2/\omega_1$.  It should be noted that for $t_d<0$ cases, the $\omega_2$ pulse precedes the $\omega_1$ pulse; conversely, the opposite is true for $t_d>0$ cases. To focus our study on the consequences of the shifting of the $\omega_2$-field, the argument of the envelope function in the second field ($\omega_2$-field) is shifted accordingly, such that only the pulse envelope shifts in time and the CEP remains unchanged.

We have considered the atomic Coulomb potential $V(r)$ for He-atom under SAE, which is modeled by an empirical expression \cite{Tong_2005}, wherein the coefficients in the empirical expression of $V(r)$ are obtained by the self-interaction free density functional theory. The ground state (initial state) energy of the He is obtained to be $\sim -0.9038$ a.u. after diagonalizing the Hamiltonian \cite{TONG1997119}.

Once the TDSE has been solved through the TDGPS method, the time-dependent dipole acceleration $\mb{a}(t)$ is assessed in accordance with the Ehrenfest theorem \cite{sandPRL_1999}:
\be
\mb{a}(t) = - \Big\langle \psi(\mb{r},t) \Big| \frac{\partial V(r)}{\partial r} + \mb{E}(t) \Big| \psi(\mb{r},t) \Big\rangle.
\ee
The harmonic spectra is then obtained by the knowledge of the Fourier transform of $\mb{a}(t)$ and is given as,
\be S(\Omega) \sim \left|\frac{a(\Omega)}{\Omega^2}\right|^2,\ee 
where $a(\Omega)$ is just the Fourier transform of the dipole acceleration. The integrated harmonic yield between the energies $E_1$ and $E_2$ are calculated as \cite{Schiessl2007_scale}, 
\be Y = \frac{1}{\tau} \int_{E_1}^{E_2} |a(\Omega)|^2 d\Omega \label{yield}\ee
where, $\tau$ is the pulse duration, $E_1$ and $E_2$ is the energy window in the plateau region.
Furthermore, the field profile of the attosecond pulse [$\mathcal{E}_{asp}(t)$] can be constructed by filtering the desired frequency range using appropriate window function $w(\Omega)$ and then taking the inverse Fourier transform as \cite{Peng-2020}:
\be \mathcal{E}_{asp}(t) = \frac{1}{\sqrt{2\pi}} \int a(\Omega) w(\Omega) e^{i\Omega t} d\Omega.\ee
We have simply used the $w(\Omega) = \Theta(\Omega - \Omega_1) \Theta(\Omega_2 - \Omega)$, wherein $\Omega_1 \leq \Omega \leq \Omega_2$ is the frequency range to be filtered, and $\Theta(x)$ is standard step function. The intensity of the attosecond pulse is then given by, \be I(t) \sim |\mathcal{E}_{asp}(t)|^2.\label{asp_calc}\ee 

\begin{figure}[b]
	\centering	\includegraphics[totalheight=0.32\textwidth]{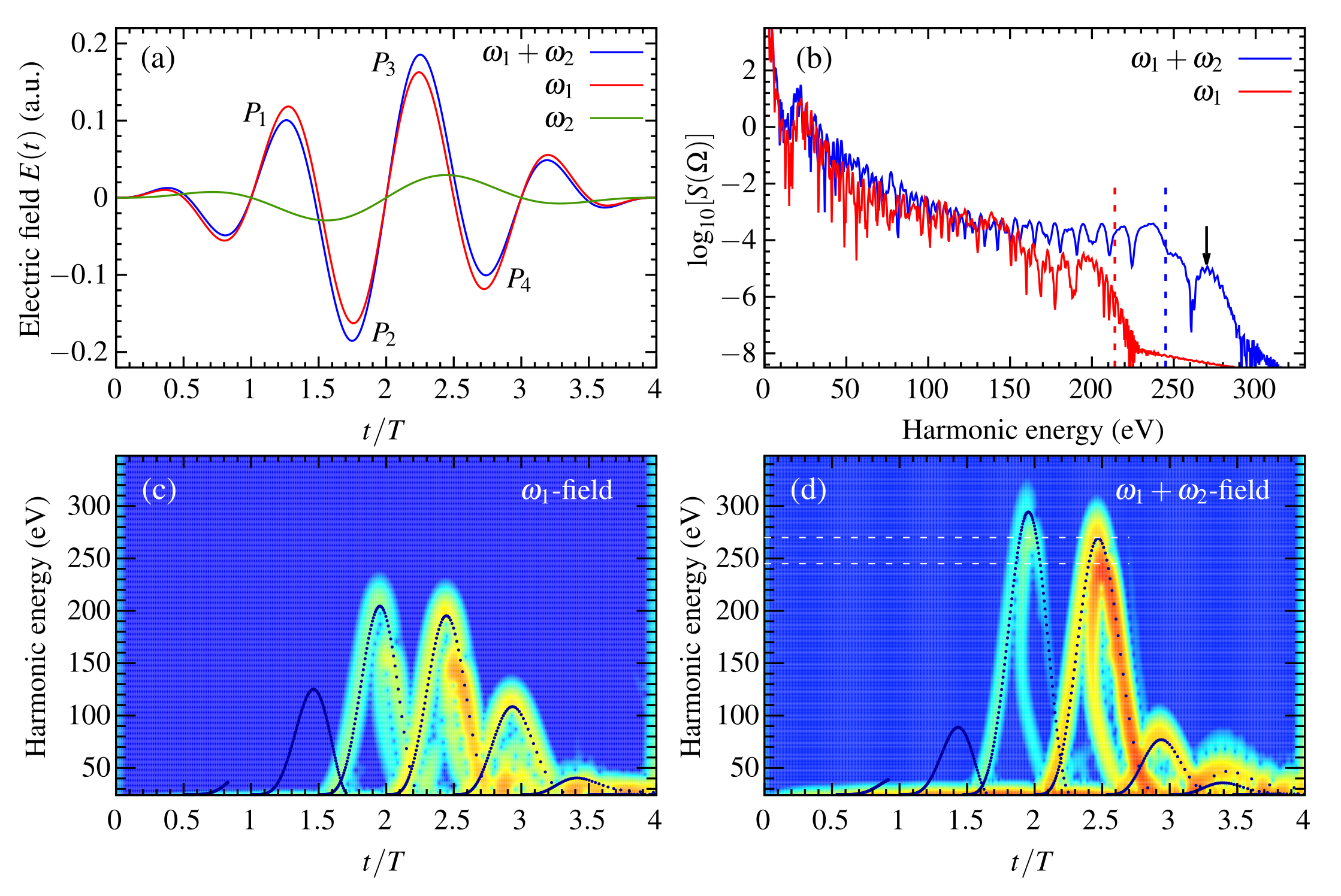}
	\caption{(a) Bi-chromatic driving field (blue) along with the constituent $\omega_1$ - field (red) and $\omega_2$ - field (green). (b) The corresponding HHG spectra are generated by the bi-chromatic driving field (blue) and the $\omega_1$-field (red) alone. The vertical dashed lines at 244 eV, and 214 eV indicate the cutoff energies $E_c = I_p + 3.17U_p$ for the bi-chromatic and $\omega_1$-field cases, respectively. While the vertical arrow at $270$ eV represents the observed cutoff energy for the bi-chromatic field. Time-frequency response of the dipole acceleration for (c) $\omega_1$-field and (d) Bi-chromatic driving field are also presented. Classical recollision energies (solid black circles) concerning the recombination times of the electron trajectories emitted for both pulses are also shown for both the driving field cases. } 
	\label{fig1}
\end{figure}
 
We considered the radial simulation domain of $r_{\text{max}} \sim 250$ a.u., and the last $30$ a.u. is used as masking to absorb the outgoing wavefunction \cite{TONG1997119}. We considered $N = 1200$ non-uniform grid points along the radial direction, $L = 60$ as maximum angular momentum, and simulation time step $0.05$ a.u. is used. The convergence is tested with respect to the spatial grid and time step. Our simulation utilizes the widely used \textit{Armadillo} library for linear algebra purposes \cite{Sanderson2016}.

\section{Results and Discussions}
\label{sec3}

First, we study the HHG for the frequency ratio $\omega_2/\omega_1 = 0.5$, followed by the study of variation in CEP and time delay on HHG. Next, the effect of the frequency ratio is explored in detail, and finally, the harmonic yield and control of attosecond pulse intensities are studied.
 
\subsection{HHG by bi-chromatic field with $\omega_2/\omega_1 = 0.5$}

\begin{figure}[b] 
	\centering	\includegraphics[totalheight=0.4\textwidth]{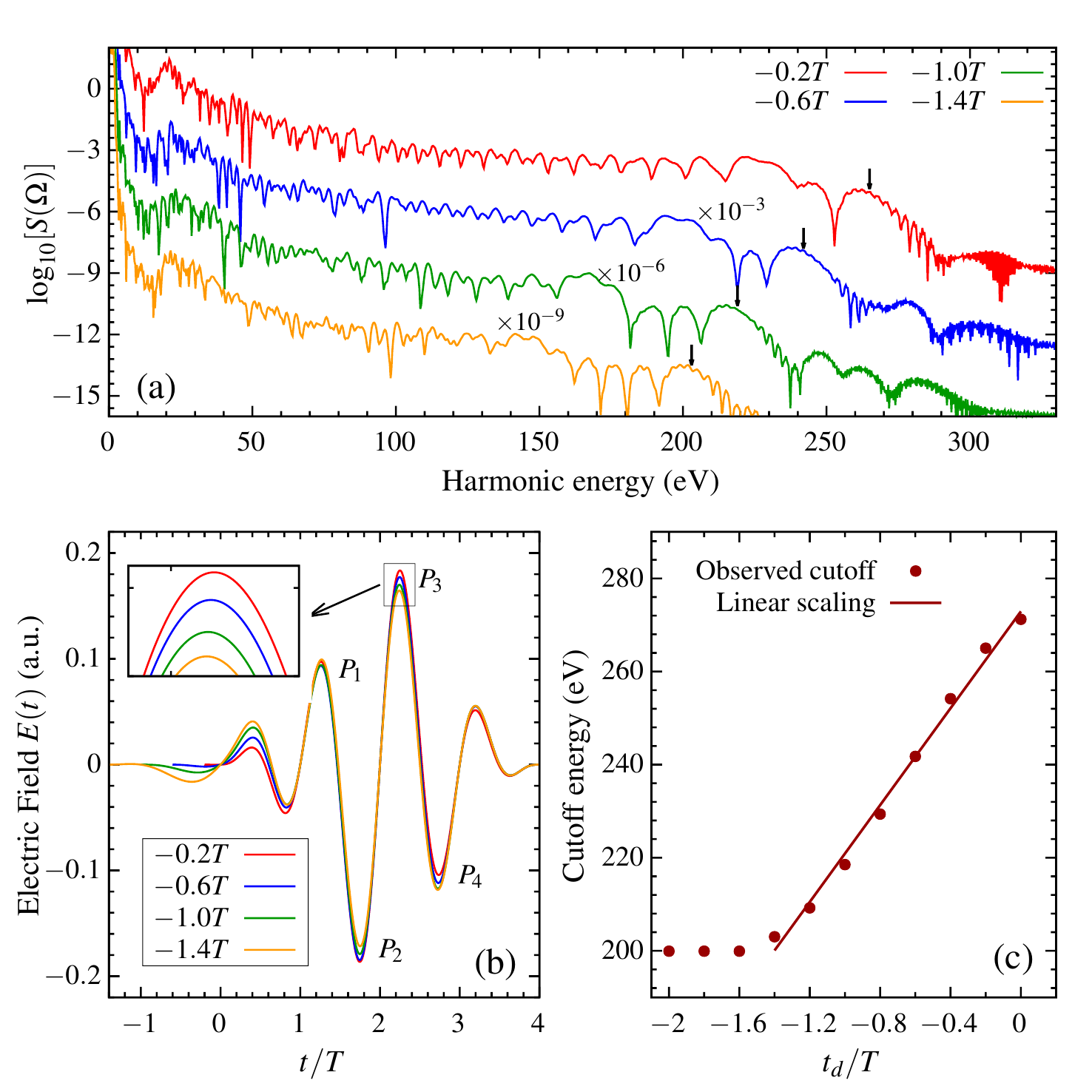}
	\caption{(a) HHG spectra generated by the bi-chromatic driving field for delay parameter $t_d$: $-0.2T$ (red), $-0.6T$ (blue), $-1.0T$ (green), and $-1.4T$ (yellow). For better visualization, the HHG spectra for $t_d = -0.6T$, $-1.0T$, and $1.4T$ are multiplied by a factor of $10^{-3}$, $10^{-6}$, and $10^{-9}$, respectively. The vertical arrows mark the position of observed cutoff energy for different delay cases. (b) The corresponding driving laser fields are shown with parameters: $\phi = 0$, $t_d \in [-1.4T, -0.2T]$, and $E_1 = 5E_2 \approx 0.17$ a.u. (c) The scaling of harmonic cutoff energy with respect to the delay parameter $t_d$ is shown. The observed cutoff energy is depicted as solid circles, while the fitted linear scaling for $-1.4T \leq t_d \leq 0T$ is shown by the solid line.} 
	\label{fig2}
\end{figure}

We present the HHG spectra of the He-atom produced by the bi-chromatic driving field [Eq. \ref{laser_field}] with parameters, $t_d = 0$ and $\phi = 0$ in Fig. \ref{fig1}. The temporal field profile of $\omega_1$ and $\omega_2$ fields are separately presented in Fig. \ref{fig1}(a), along with the resultant bi-chromatic driving field $\omega_1 + \omega_2$. Two major ionization peaks corresponding to both bi-chromatic ($\omega_1 + \omega_2$) and primary $\omega_1$-field are also shown in Fig. \ref{fig1}(a). The electrons are ionized from the peak marked as $P_1$ and, subsequently, return to emit the HHG radiation between the peaks $P_2$ and $P_3$. Similarly, the electrons ionizing due to peak $P_2$ are returning between the peaks $P_3$ and $P_4$ to produce the HHG radiation. The corresponding HHG spectra generated by the bi-chromatic driving field and the primary $\omega_1$-field alone are compared in Fig. \ref{fig1}(b). The harmonic spectrum (red curve) of the $\omega_1$-field alone exhibits a plateau with the cutoff position at an energy $\sim 214$ eV (marked by the red dashed line), which agrees well with the cutoff formula $I_p + 3.17U_p$, where $I_p = 24.6$ eV is the first ionization energy of the He-atom and $U_p = E_1^2/(4\omega_1^2)$ is the ponderomotive energy corresponding to the $\omega_1$-field. For the bi-chromatic driving field (black curve) the harmonic cutoff is extended significantly to $\sim 270$ eV, as shown in Fig. \ref{fig1}(b). For the case of the bi-chromatic field, the cutoff formula $I_p + 3.17U_p'$, where $4U_p' = E_1^2/\omega_1^2 + E_2^2/\omega_2^2$, predict the cutoff energy $\sim 245$ eV (marked by the black dashed line). However, we observed a secondary cutoff $\sim 270$ eV (marked by the vertical arrow) in the bi-chromatic field case. This secondary cutoff is well beyond the predicted value (i.e., $245$ eV) and corresponds to the returning kinetic energy of $3.54U_p'$. Furthermore, the intensity of the harmonics near the cutoff is low compared to the harmonics in the plateau region.

\begin{figure}[t] 
	\centering	\includegraphics[totalheight=0.4\textwidth]{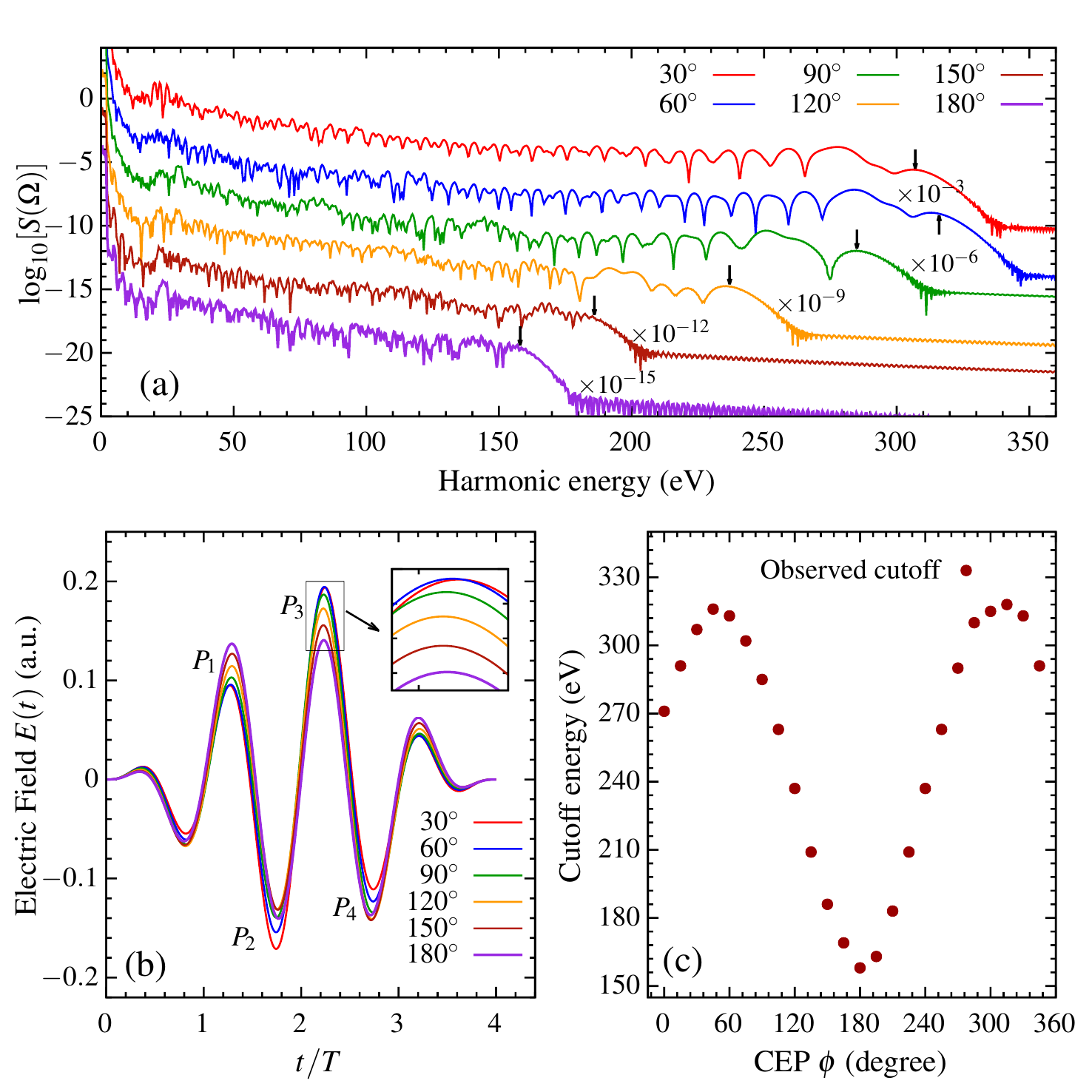}
	\caption{(a) HHG spectra generated by the bi-chromatic driving field for the CEP $\phi$: $30\degree$ (red), $60\degree$ (blue), $90\degree$ (green), $120\degree$ (yellow), $150\degree$ (maroon), and $180\degree$ (violet). For better visualization, the HHG spectra in the range $\phi \in [60\degree, 180\degree]$ are multiplied by the factors shown in the figure. The vertical arrows represent the position of observed cutoff energy for the individual cases. (b) The corresponding driving laser fields are shown with parameters: $t_d = 0$, $\phi \in [30\degree, 180\degree]$, and $E_1 = 5E_2 \approx 0.17$ a.u. (c) The harmonic cutoff energy concerning the CEP $\phi$ is shown for $0\degree \leq \phi \leq 360\degree$.} 
	\label{fig3}
\end{figure}

In order to understand the dynamics that lead to the observed structure of the HHG spectra, the time-frequency analysis is carried out using the Gabor transformation. Additionally, the classical recollision energies of the electron trajectories emitted for the bi-chromatic and $\omega_1$-field alone are also determined. In Fig. \ref{fig1}(c,d), the time-frequency analysis along with the classical recollision energies (solid black circles) is shown. The classical recollision energies are up-shifted by the ionization potential of the He atom. It can be seen from Fig. \ref{fig1}(c) ($\omega_1$-field case), that the harmonic emission is taking place after $t = 1.5T$ which peaks around $\sim 214$ eV, corroborating the harmonic cutoff observed in Fig. \ref{fig1}(b) [for $\omega_1$-field case (red curve)]. Moreover, there are two prominent emission peaks contributing to the HHG. The interference of these multiple emission events leads to the modulations in HHG spectra. The results of time-frequency analysis are further validated by the classical recollision energies (solid black circles). However, an emission peak is present in classical trajectory analysis around $t = 1.5T$, which is originated due to the recombination of electrons ionized around $t=0.7T$ and is missing in the time-frequency response. The plausible reason for this disappearance can be given with the help of Fig. \ref{fig1}(a), wherein the temporal variation in driving field amplitude is shown. It can be observed that the amplitude of the laser field for $t<0.7T$ is indeed tiny. Consequently, the ionization probability is insignificant, and the subsequent emission of harmonics can be ignored, which is in agreement with the time-frequency response shown in Fig. \ref{fig1}(c). The time-frequency analysis of the HHG spectra corresponding to the bi-chromatic driver is presented in Fig. \ref{fig1}(d). It can be observed that the emission contributing to the HHG is also originating from $t= 1.5T$ in the bi-chromatic driver case, and similar to the $\omega_1$ - field case, two main emission peaks are present before the cutoff energy $\sim 270$ eV. Additionally, the intensity of the emitted harmonics is relatively low in the energy range $245 - 270$ eV, which can also be observed in the HHG spectra (blue) depicted in Fig. \ref{fig1}(b). 

\subsection{Effect of pulse delay and CEP on the HHG}

\begin{figure}[b] 
	\centering	\includegraphics[totalheight=0.32\textwidth]{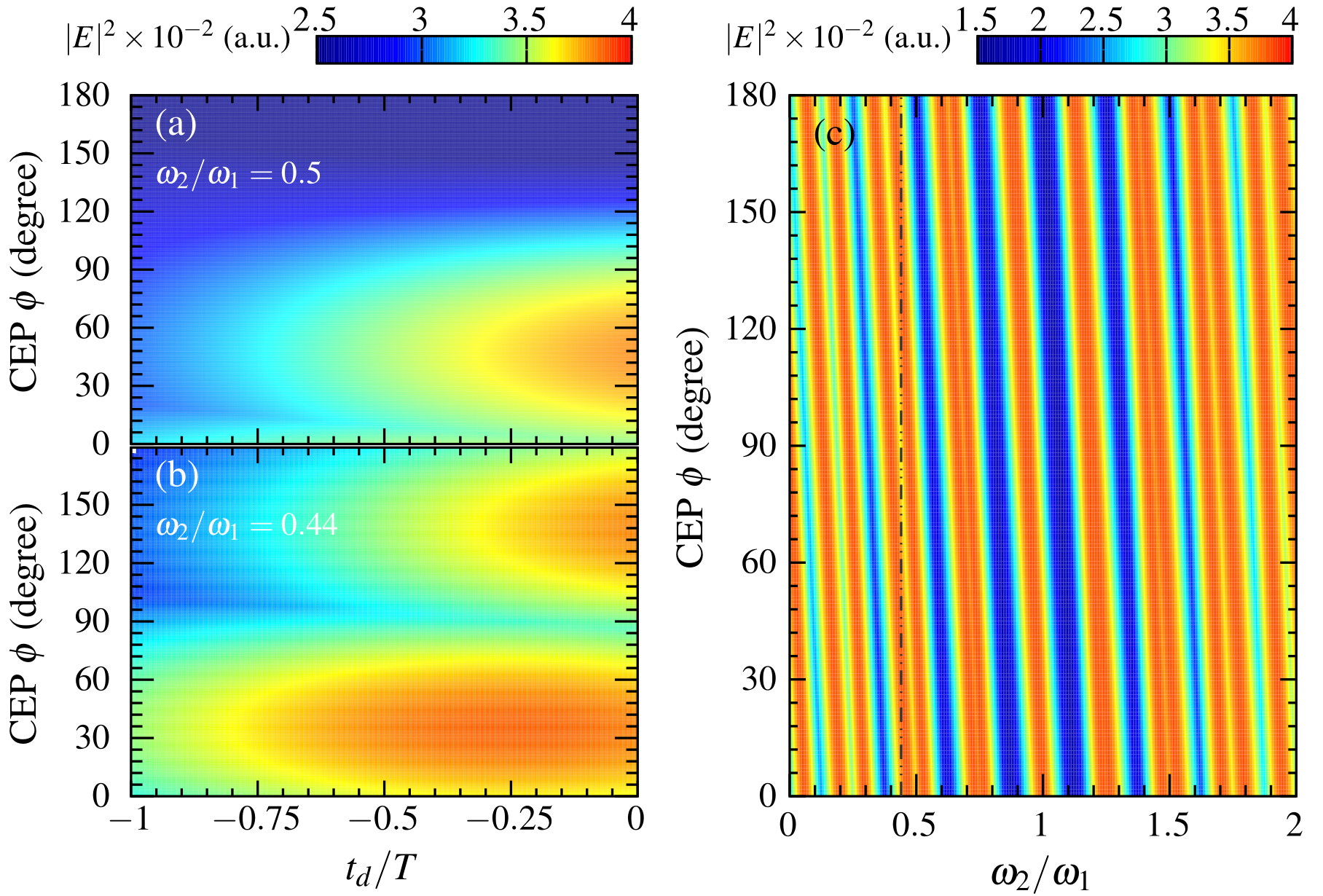}
	\caption{Variation of the peak field intensity $\sim |E(t)|^2$ with CEP phase $\phi$ and delay parameter $t_d$ is presented in  fixed frequency ratio $\omega_2/\omega_1 = 0.5$ (a) and $\omega_2/\omega_1 = 0.44$ (b). However, the variation of peak intensity with the frequency ratio and CEP phase for fixed time delay $t_d = 0$ is presented in (c). A vertical dashed line in (c) highlights $\omega_2/\omega_1 = 0.44$ value.}
\label{fig4}
\end{figure}
 
Next, we investigate the effect of the time delay between the two pulses on HHG spectra. In Fig. \ref{fig2}(a), the harmonic spectra generated by the bi-chromatic field for different delay duration are shown. We considered the delay cases for $t_d<0$, which means that the $\omega_2$-field precedes the $\omega_1$-field. The HHG spectra corresponding to delay parameter $t_d \le -0.6T$ are scaled appropriately [refer figure caption]. The harmonic cutoff energy is marked with a vertical arrow for each delay case. It can be observed that the cutoff energy varies smoothly with the delay parameter because the intensity of the local maximum of the corresponding driving fields also varies, as shown in Fig. \ref{fig2}(b). It can be noticed that the variation in delay parameter $t_d$ results in the variation of the intensity of peak $P_3$. The intensity of peak $P_3$ is maximum for $t_d = -0.2T$ and decreases gradually as $t_d$ becomes more and more negative [see inset of Fig. \ref{fig2}(b)]. This implies that the energy gained by the electron decreases with an increasing delay between the two combining pulses. As a result, the associated cutoff energy of the respective HHG spectra scales linearly. In Fig. \ref{fig2}(c), the observed cutoff energy (solid maroon circles) for different values of delay parameter ($t_d$) are presented. It is observed that the cutoff energy scales linearly with $t_d$ in the range $-1.4T < t_d < 0T$. This scaling of cutoff energy is mainly because of the variation in the electron ponderomotive energy with varying strength of the driving electric field. For $t_d< -1.4T$ cases, the strength of the $\omega_2$-field is noticeably reduced near the peak $P_3$. As a result, there is no significant change occurs in the intensity of the peak $P_3$. Therefore, the position of the observed cutoff energy remains unchanged for $t_d < -1.4T$ cases.

We have also investigated the effect of the CEP parameter $\phi$ on the HHG cutoff energy. In Fig. \ref{fig3}(a), the HHG spectra for the range $\phi \in [30\degree,180\degree]$ are presented. The harmonic cutoff energy for $\phi=30\degree$ is observed at $\sim 307$ eV and increases to $\sim 313$ eV for $\phi=60\degree$ case. Further increase in the CEP, $\phi$, results in a decrease in the cutoff energy of HHG spectra. This variation in energy can be understood by examining the corresponding driving laser fields. In Fig. \ref{fig3}(b), the driving laser fields for different $\phi$ values are shown. As can be seen that the variation in the parameter $\phi$ causes the variation of the intensity of peak marked as $P_3$ in Fig. \ref{fig3}(b). The amplitude of peak $P_3$ increases as $\phi$ varies from $30\degree$ to $60\degree$ and then gradually decreases for increasing values of $\phi$ [see inset of Fig. \ref{fig3}(b)]. This modulation in the strength of laser pulse peak $P_3$ translates to the ponderomotive energy of the electron and the subsequent cutoff energy of the emitted harmonic radiation changes. Also, we have calculated the cutoff energy for an entire range of CEP, $\phi$, between $0\degree$ to $360\degree$ and presented in Fig. \ref{fig3}(c). The cutoff energy oscillates between the values $\sim 158$ eV and $\sim316$ eV. It is maximum for $\phi = 45\degree$ and $325\degree$ cases, while decreases smoothly for any variation in $\phi$ on either side of these two angles and is minimum at $\phi=180\degree$.

\subsection{Effect of different frequency ratio}

So far we explored the HHG using a bi-chromatic laser pulse by keeping $\omega_2 = 0.5 \omega_1$. We observed that the relative phase $\phi$ of the two pulses and the time delay between the two pulses $t_d$ were very instrumental in controlling the quantum dynamics of the electron and so the HHG as a process. Let's study how the frequency ratio $\omega_2/\omega_1$ affects the harmonic emission process. As we know, the peak intensity of the pulse $\sim |E(t)|^2$ directly affects the electron's pondermotive energy ($U_p$). Hence, the knowledge of the peak intensity with the CEP phase, time delay, and frequency ratio can shed light on the crucial aspects of the harmonic emission process. In order to elucidate the effect of the relative phase ($\phi$) and the time delay ($t_d$) between the two pulses, in Fig. \ref{fig4}(a,b) we present the variation of the maximum intensity $\sim |E(t)|^2$ of the pulse as a function of the $\phi$ and $t_d$ for a fixed ratio $\omega_2/\omega_1 = 0.5$ and $\omega_2/\omega_1 = 0.44$ by using Eq. \ref{laser_field}. As can be seen from this figure, the maximum peak intensity occur for $t_d = 0$ and $\phi = 45^\circ$ in case of $\omega_2/\omega_1 = 0.5$ (a) and $\phi = 30^\circ$ and $150^\circ$ for $\omega_2/\omega_1 = 0.44$ case (b). It can be seen from Fig. \ref{fig4}(a) that for $\phi = 180^\circ$ the peak intensity is found to be minimum, as the primary and the second pulse would be superposed out of phase for frequency ratio $\omega_2/\omega_1 = 0.5$. This feature of the CEP phase is also corroborated in Fig. \ref{fig3}(a) [where we used $\omega_2 = 0.5\omega_1$]. Furthermore, the variation of the peak amplitude with the delay parameter is very apparent from Figs. \ref{fig4}(a) and (b). The time delay between the two pulses is quite feasible to control from an experimental perspective. Hence, it can be used to control the electron dynamics to favor enhanced harmonic emission. 
 
 \begin{figure}[t] 
	\centering	\includegraphics[totalheight=0.3\textwidth]{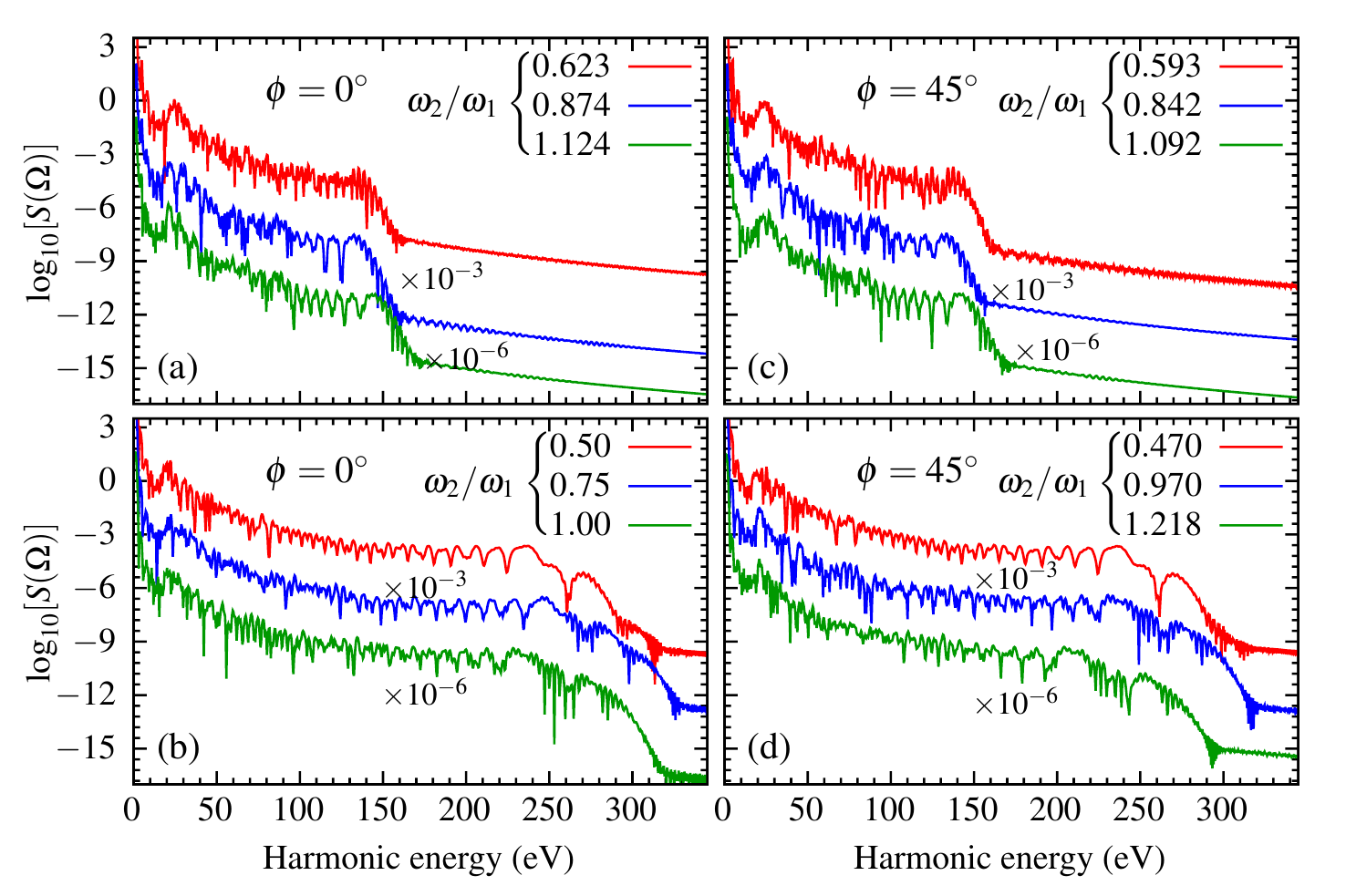}
	\caption{The harmonic spectra for different frequency ratio $\omega_2/\omega_1$ for $\phi = 0^\circ$ (a,b) and $\phi = 45^\circ$ (c,d) are presented. Here, we have considered $t_d = 0$. The curves in each sub-figures are shifted downward by the factor mentioned in the plots.}
\label{fig5}
\end{figure}
 
In Fig. \ref{fig4}(c), we present the maximum peak intensity of the pulse as a function of the frequency ratio $\omega_2/\omega_1$ and CEP phase $\phi$, for a fixed delay parameter $t_d = 0$. As we note from Fig. \ref{fig4}(c), there are alternate bright and dark fringes as we vary the ratio $\omega_2/\omega_1$ with $\phi$. The dark fringes represent smaller peak intensity, and the bright fringes represent a more significant peak intensity. Moreover, the fringes are found to be slightly tilted toward a lower ratio as we increase the CEP phase, which is expected from the superposition principle; as the CEP changes, the electric field of primary and secondary pulse go out of phase, causing out of phase addition of the two profiles and lowering the intensity. A vertical line in Fig. \ref{fig4}(c) represents $\omega_2/\omega_1 = 0.44$ value; it can be seen that this frequency ratio would yield maximum peak intensity at two different CEP phase values, which we also observed in Fig. \ref{fig4}(b). 

\begin{figure}[t] 
	\centering	\includegraphics[totalheight=0.32\textwidth]{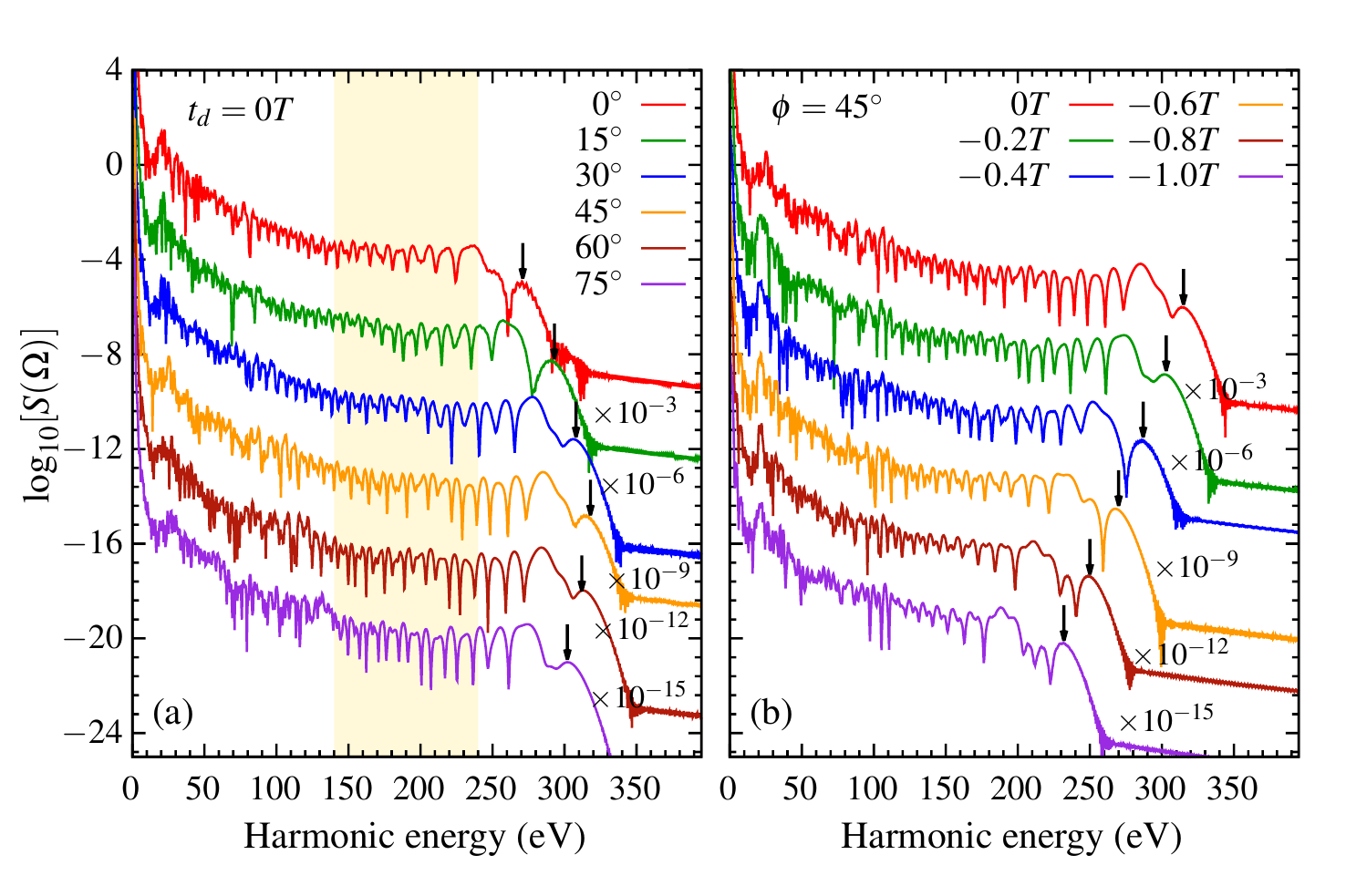}
	\caption{The HHG spectra for different CEP phase $\phi$ is presented for fixed $t_d = 0$ (a), and for fixed CEP $\phi = 45^\circ$, the high harmonic spectra are shown by varying time delay between two pulses (b). The frequency ratio is considered to be $\omega_2 = 0.5\omega_1$. The highlighted portion of (a) is the energy range 140 eV $-$ 240 eV, which is later used to construct the attosecond pulses in Fig. \ref{fig7}.}
\label{fig6}
\end{figure}

In order to understand how the choice of the frequency ratio $\omega_2/\omega_1$ affects the harmonic emission, we have compared the HHG spectra for $\phi = 0^\circ$ and $\phi = 45^\circ$ with $t_d = 0$ in Fig. \ref{fig5}. The respective frequency ratios are mentioned in the figure legends. The plots presented in Fig. \ref{fig5}(a) and (c) for the mentioned frequency ratio belong to the dark fringes in Fig. \ref{fig4}(c). However, the plots in Fig. \ref{fig5}(b) and (d) are associated with the frequency ratios belonging to bright fringes in Fig. \ref{fig4}(c) for respective CEP phases. As can be seen, the harmonic cutoff in Fig. \ref{fig5}(a) and (c) is $\sim 150$ eV, and in Fig. \ref{fig5}(b) and (d) is $\sim 270$ eV or so, which is mainly because of the variation in the peak intensity of the driving bi-chromatic laser pulse. We have covered a vast range of frequency ratios from $\omega_2/\omega_1 < 1$ and $\omega_2/\omega_1 > 1$, and in all cases, the CEP phase would dictate the maximum peak intensity of the laser pulse and so the harmonic cutoff observed. From Fig. \ref{fig4} and Fig. \ref{fig5} we can say that depending on the experimental availability of the secondary laser pulse ($\omega_2$-field), an appropriate CEP phase can be chosen to optimize the harmonic emission for maximum cutoff energy. Alternatively, the frequency ratio of the bi-chromatic driver field can also be tuned accordingly for enhanced harmonic cutoff, as lately, some experimental research groups have proposed tuning the laser wavelength using an optical parametric amplifier \cite{Musheghyan_2020,HONG2018169}. It is interesting to note that increasing the frequency ratio $\omega_2/\omega_1$ does not guarantee that the harmonic cutoff can be enhanced, however, the relative CEP between the two pulses plays a crucial role in determining optimum harmonic emission. 

We observed in Fig. \ref{fig4}(a) that for a given frequency ratio, the peak intensity of the bi-chromatic driving field changes with the CEP $\phi$ and the delay between the two pulses $t_d$. In order to highlight the effect of this variation, in Fig. \ref{fig6} we have presented the HHG spectra for $\omega_2/\omega_1 = 0.5$ with variation in $\phi$ [Fig. \ref{fig6}(a)] and time delay $t_d$ [Fig. \ref{fig6}(b)]. We can see from Fig. \ref{fig6}(a) that with $t_d = 0$, the cutoff energy is found to be maximum for $\phi = 45^\circ$. However, for a fixed $\phi = 45^\circ$, the harmonic cutoff decreases with increasing the delay [Fig. \ref{fig6}(b)]. The motivation of presenting Fig. \ref{fig6} [apart from Fig. \ref{fig2}(a) and Fig. \ref{fig3}(a)] is to demonstrate the possibility of scanning the parameter space along vertical or horizontal direction of Fig. \ref{fig4}(a), and in either case a very fine control over the harmonic cutoff energy is observed in conjunction with Fig. \ref{fig4}. Furthermore, the highlighted portion in Fig. \ref{fig6}(a) shows the energy range 140 eV $-$ 240 eV, which we have used to study the harmonic yield and associated attosecond pulses as discussed next and presented in Fig. \ref{fig7}. 
    
\subsection{Harmonic yield and attosecond pulse generation} 

\begin{figure}[t] 
	\centering	\includegraphics[totalheight=0.32\textwidth]{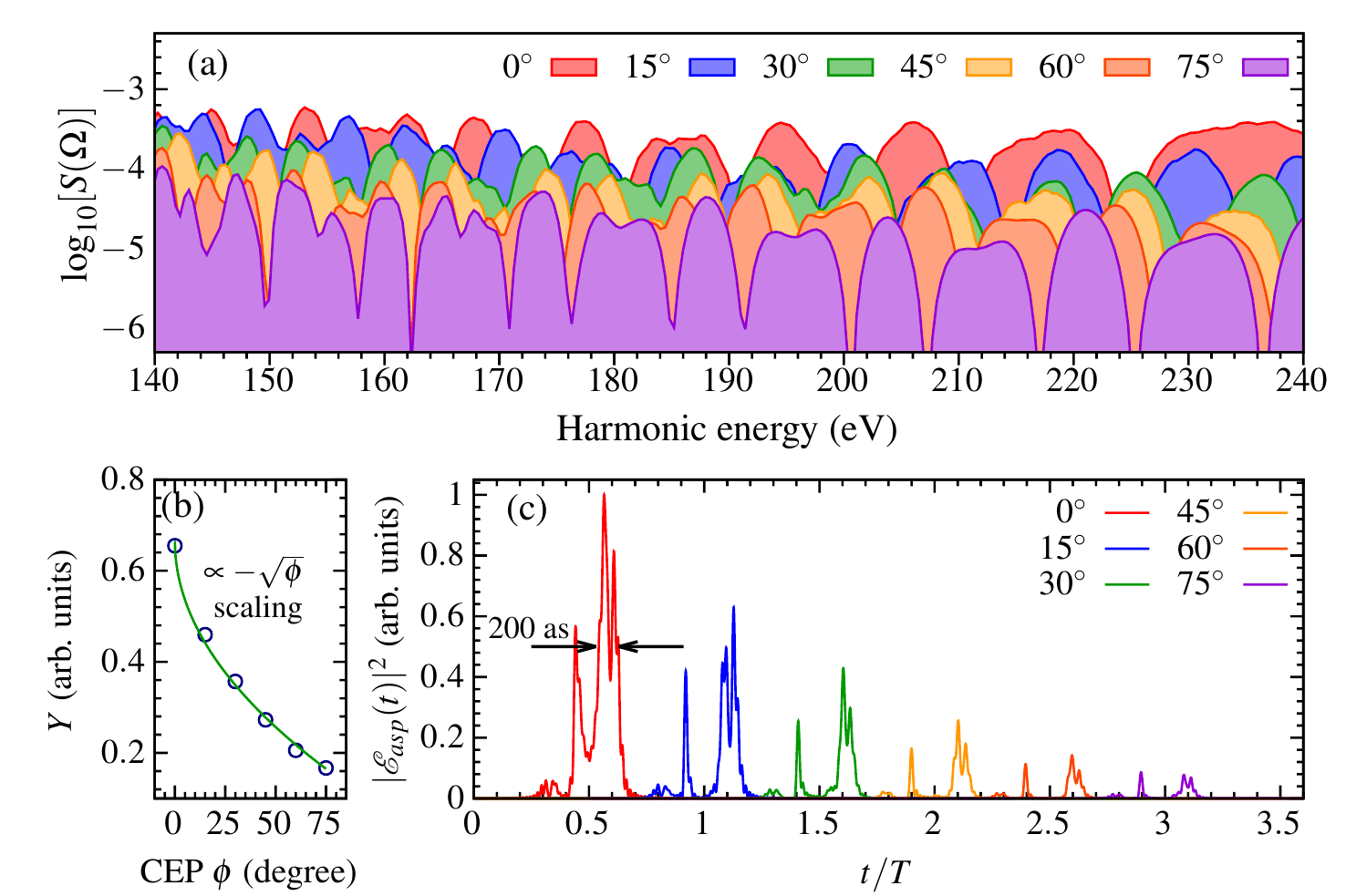}
	\caption{The HHG spectra in the energy range 140 eV $-$ 240 eV for different CEP phases is presented in (a), which is the highlighted portion of Fig. \ref{fig6}(a) without any scaling factor. The harmonic yield in the mentioned energy range is calculated as a function of CEP phase $\phi$ and presented in (b). In (b), the open circles show the observed yield from the simulation, and the solid line shows the $Y \propto - \sqrt{\phi}$ scaling. The attosecond pulse for different $\phi$ is illustrated in (c). The peak of the attosecond pulses is normalized with respect to the $\phi = 0^\circ$ case, and attosecond pulses are intentionally shifted by $0.5T$ for better representation. } 
\label{fig7}
\end{figure}

Now we consider the attosecond pulse generation for $\omega_2 =0.5\omega_1$ case with $t_d = 0$. Figure \ref{fig6}(a) illustrates the harmonic spectra with variation in CEP. We filtered the harmonics in the range 140 eV $-$ 240 eV [highlighted portion in Fig. \ref{fig6}(a)] and calculated the respective harmonic yield [refer Eq. \ref{yield}]. In Fig. \ref{fig7}(a), we present the filtered harmonic spectra [without any scaling factor] for different $\phi$ cases, and it can be seen that harmonic efficiency decreases with increasing $\phi$. The harmonic yield for all the cases is also shown in Fig. \ref{fig7}(b) with open circles. The harmonic yield is observed to follow $\propto -\sqrt{\phi}$ scaling with CEP to good accuracy. The attosecond pulses are constructed [Fig. \ref{fig7}(c)] by superposing the harmonics in the range 140 eV $-$ 240 eV [refer Eq. \ref{asp_calc}]. The peak intensity of the attosecond pulses is normalized with respect to the $\phi = 0^\circ$ case, and each pulse is shifted by $0.5T$ for better visual representation. We are superposing the same harmonics. Hence it is observed that the attosecond pulse duration for all the cases is $\sim 200$ as; however, the peak intensity of the same can be tweaked in a controlled way by varying the CEP $\phi$. The control of the attosecond pulses is crucial as the `Attosecond Science' is quite an emerging field, promising vast applications in both fundamental and applied sciences \cite{Calegari_2016,Guo_2018,Yang2021}.

\section{Summary and Conclusions}
\label{sec4}

We studied the high-order harmonic generation by a time-delayed, linearly polarized bi-chromatic laser [Eq. \ref{laser_field}] pulse with Helium target. The time delay between the two pulses, their frequency ratio, and the relative CEP plays a very prominent role in determining the ionization of the electron and associated electron dynamics in the continuum, resulting in a high harmonic generation. We studied the cases with $t_d < 0$, which implies that the $\omega_2$-field precedes the fundamental $\omega_1$-field. It has been observed that in comparison with the $\omega_1$-field alone, the maximum emitted harmonic energy is significantly increased for the bi-chromatic laser pulse. We also observed that the energy of the cutoff harmonic varies linearly with the delay between the two component pulses in the range $-1.4T \leq t_d \leq 0T$. The cutoff energy is also seen to vary in a periodic manner with the CEP $\phi$, wherein the bi-chromatic driver has no time delay between the component pulses. 

In order to explore the effect of the frequency ratio, we have calculated the variation of the peak laser intensity as a function of the frequency ratio $\omega_2/\omega_1$, CEP phase $\phi$ and the time delay between the two pulses $t_d$. It is observed that there is an optimum CEP phase for a given frequency ratio and $t_d$. An extensive set of parameters in $\phi$, $\omega_2/\omega_1$ and $t_d$ are scanned, and it is observed that the relative CEP is equally crucial for a given frequency ratio to have optimum harmonic cutoff energy. The harmonic yield in the energy range 140 eV $-$ 240 eV is also computed for the ratio $\omega_2/\omega_1 = 0.5$, and it is found that the harmonic yield smoothly scales as $\propto -\sqrt{\phi}$. As a result, the intensity of generated attosecond pulses of duration $\sim 200$ as is controlled in a predictable manner. The advantage of the scheme used here lies in the tuning of the cutoff energy of the emitted harmonics by merely adjusting the delay between the component pulses and varying the CEP, which is feasible from an experimental point of view.

In this work, we relied on the single atom response, but in the actual scenario, the macroscopic propagation effects play a crucial role \cite{Rivas2018,ALHuillier_1991}. A numerical model to address the macroscopic propagation of the emitted harmonics take the induced dipole moment as an input and then evolves the same using wave propagation equations \cite{priori2000_PRA,Heyl_2017,PhysRevA.99.033411}. In our future work, we intend to incorporate the macroscopic propagation effects, which are currently out of the scope of the present manuscript. 
 
\section*{Acknowledgments} Authors would like to acknowledge the DST-SERB, Government of India, for funding the project CRG/2020/001020. 
 
\bibliographystyle{apsrev4-2}

%

\end{document}